\documentclass [12pt,epsfig,showkeys]{article} 

\def\be{\begin{equation}} \def\ee{\end{equation}} 
\def\bea{\begin{eqnarray}}
\def\eea{\end{eqnarray}} 

\def\mN{m_{\mbox{\tiny N}}}

\begin{document}
\renewcommand{\thefootnote}{\fnsymbol{footnote}}
\setcounter{footnote}{1}


\begin{center}
\vskip 5mm 
{\Large\bf 
Comparison of the extended linear sigma model 
and chiral perturbation theory }
\vskip 5mm 
{\large 
W.P. Alvarez$^{(a,b)}$\footnote{E-mail:alvarezwilson@hotmail.com},
K. Kubodera$^{(b)}$\footnote{E-mail:kubodera@sc.edu},
F. Myhrer$^{(b)}$\footnote{E-mail:myhrer@sc.edu},
} 

\vskip 5mm

{\it 
${}^{(a)}$ 
Facultad de Ingener\'{i}a, Ciencias F\'{i}sicas 
y Matem\'{a}ticas, 
Universidad Central del Ecuador, 
Quito,Ecuador\\
${}^{(b)}$Department of Physics and Astronomy,
University of South Carolina,
Columbia, \\
SC 29208, USA } 
\end{center}

\vskip 5mm

The pion-nucleon scattering amplitudes 
are calculated in tree approximation
with the use of the extended linear sigma model
(ELSM) as well as heavy baryon chiral perturbation 
theory (HB$\chi$PT),
and the non-relativistic forms 
of the ELSM results 
are compared with those of HB$\chi$PT.
We find that the amplitudes obtained in ELSM
do not agree with those derived from 
the more fundamental effective approach,
HB$\chi$PT.

\newpage
\renewcommand{\thefootnote}{\arabic{footnote}}
\setcounter{footnote}{0} 

The linear sigma model 
\cite{GML60},
which provides an illuminating example 
of spontaneous chiral 
symmetry breaking in strong interactions,
has been studied extensively in the literature.
Some of the consequences of this model, 
however, are known to be in conflict with observation.
Notably, 
the isoscalar pion-nucleon ($\pi N$) scattering length
predicted by the model 
is larger than the experimental value
by an order of magnitude.  
Furthermore, the model predicts 
the axial coupling constant $g_A$ to be unity,
whereas empirically $g_A\approx 1.26$. 
Despite the known limitations of the model,
its simplicity has invited many authors
to use it for exploring the consequences of 
chiral symmetry in nuclear physics;
see e.g. Ref.\cite{ericson}. 
Nauenberg and Bjorken \cite{bjn68}
and Lee \cite{lee68} 
introduced an extended linear sigma model
(ELSM) by adding a pair of extra terms 
(which jointly preserve chiral symmetry)
to the original linear sigma model lagrangian. 
An important feature of ELSM 
is that $g_A$ is no longer restricted to be unity.
Furthermore, via chiral rotations of the fields, 
ELSM leads to the non-linear chiral lagrangian
of Weinberg (with $g_A \ne 1$) in 
the limit of an infinitely massive scalar field.
Recently ELSM has been used to investigate
the $g_A$ dependence of the 
$\pi N$ scattering lengths and the 
$\pi N$ sigma term $\Sigma_N$ \cite{dm00}. 
It has been found in Ref.\cite{dm00}
that ELSM can reproduce
the very small experimental value of the $\pi N$ 
isoscalar scattering length, $a_{\pi N}^{(+)}$, 
and furthermore the same model can reproduce
the large empirical value of  the 
$\pi N$ sigma term, $\Sigma_N$, 
without invoking any $\bar{s} s$ component 
of the nucleon. 

\vspace{3mm}

Meanwhile, low-energy hadronic physics 
can be described by an effective field theory 
(EFT) of QCD known as ``chiral perturbation theory'' 
($\chi$PT)~\cite{weinberg,wei90}.
The $\chi$PT Lagrangian, ${\cal L}_{\chi PT}$,
reflects the symmetries 
and the pattern of symmetry breaking 
of the underlying QCD. 
${\cal L}_{\chi PT}$ is expanded in powers of
$Q/\Lambda_\chi \ll 1$ 
where $Q$ denotes the typical four-momentum  
of the process in question or 
the pion mass, $m_\pi$, which represents the small explicit 
chiral symmetry breaking scale; 
$\Lambda_\chi \simeq 4\pi f_\pi \simeq $ 1 GeV,   
is the chiral scale.
The parameters appearing in ${\cal L}_{\chi PT}$,
called the {\it low-energy constants} (LEC's),
effectively subsume the high-energy physics 
that has been integrated out.
These LEC's could in principle be determined 
from the underlying theory,   
but in practice they are 
fixed phenomenologically from 
experimental data. 
Once the LEC's are determined, 
${\cal L}_{\chi PT}$ represents a complete and
hence model-independent Lagrangian up to a 
specified chiral order.
Furthermore, starting from ${\cal L}_\chi$,
one can develop, for the amplitude 
of a given process,
a well-defined perturbation scheme
by organizing the relevant Feynman diagrams 
according to powers in 
$Q/\Lambda_\chi$. 
If all the Feynman diagrams up to a given power, $\nu$, 
in $Q/\Lambda_\chi$ are taken into account, 
then the results are model-independent up to
this order, with the contributions of higher order terms 
suppressed by an extra power of $Q/\Lambda_\chi$.
A problem one encounters in extending $\chi$PT
to the nucleon sector is that,
as the nucleon mass $\mN$ is comparable to 
the cut-off scale $\Lambda_{\chi}$, 
a straightforward application of expansion in $Q/\Lambda$
becomes difficult~\cite{gss88}. 
This difficulty can be circumvented by employing 
heavy-baryon chiral perturbation theory (HB$\chi$PT)
\cite{jm91},
which essentially consists in shifting 
the reference point of the nucleon energy 
from 0 to $\mN$ and in integrating out 
the small component of the nucleon field 
as well as the anti-nucleonic degrees of freedom.
An effective Lagrangian in HB$\chi$PT therefore involves 
as explicit degrees of freedom
the pions and the large components 
of the redefined nucleon field.
The expansion parameters in HB$\chi$PT are
$Q/\Lambda_{\chi}$, $m_\pi/\Lambda_{\chi}$
and $Q/\mN$.
Since $\mN\approx \Lambda_{\chi}$,
it is convenient to combine chiral 
and heavy-baryon expansions
and introduce the chiral index ${\bar \nu}$ 
defined by $\bar{\nu}=d+(n/2)-2$.
Here $n$ is the number of fermion lines 
that participate in a given vertex,
and $d$ is the number of derivatives
(with $m_\pi$ counted as one derivative). 
A similar power counting scheme can also
be introduced
for Feynman diagrams as well~\cite{wei90,vanK99}.
HB$\chi$PT has been used with great success
to the one-nucleon sector, 
see, {\it e.g.}, Ref.~\cite{bkm95}. 

\vspace{3mm}

We therefore consider it informative 
to compare the predictions 
of ELSM with those of HB$\chi$PT\cite{comment1}.
As an example of this comparison,
we consider here 
the tree-level $\pi N$ scattering amplitudes 
calculated in ELSM and HB$\chi$PT to lowest order corrections in 
$Q/\Lambda_\chi$.

\vspace{3mm}

The lagrangian of the extended sigma model 
(ELSM) consists of the 
standard linear sigma model lagrangian 
plus two pion-nucleon interaction terms 
with a common coupling constant 
proportional to $(g_A - 1)$. 
The additional terms are 
a vector- 
and a pseudo-vector coupling term \cite{bjn68,lee68}. 
Thus the lagrangian of ELSM reads  
\bea
{\cal L} &=& \bar\psi i \partial \psi 
- g \bar \psi \Big[\sigma + 
i \gamma_5 \vec{\pi}\cdot \vec{\tau} \Big] 
\psi +\frac{1}{2} \Big[ (\partial_\mu \sigma)^2 + 
(\partial_\mu \vec{\pi})^2 \Big] 
\nonumber \\ & & 
+\frac{1}{2} \mu_0^2\Big[\sigma^2+\vec{\pi}^2\Big] 
-\frac{\lambda}{4}\Big[\sigma^2+\vec{\pi}^2\Big]^2 
+{\cal L}_{\chi sb} 
\nonumber \\ &&
+\left(\frac{g_A-1}{f_\pi^2}\right) 
\Big[ 
\left(\bar\psi \gamma_\mu \frac{\vec{\tau}}{2}\psi
\right) 
\cdot \left(\vec{\pi}\times \partial^\mu \vec{\pi}
\right) 
+ \left(\bar\psi\gamma_\mu\gamma_5 
\frac{\vec{\tau}}{2}\psi\right) 
\cdot\left(\sigma \partial^\mu\vec{\pi} 
-\vec{\pi}\partial^\mu \sigma\right)
\Big]\,,
\label{eq:Lsigma}
\eea
where the parameters $\lambda$ and $\mu_0$ 
are assumed to be real and positive. 
The last line proportional to $(g_A-1)$ 
represents the additional $\pi N$ coupling terms 
introduced in \cite{bjn68,lee68}. 
As for the explicit chiral symmetry breaking term
${\cal L}_{\chi sb}$,
we consider three terms
(see e.g., Refs.\cite{c79,bc79}): 
\bea
{\cal L}_{\chi sb} &=& \varepsilon_1 \sigma 
-\varepsilon_2 \vec{\pi}^2 -\varepsilon_3 
\bar\psi \psi 
\label{eq:Lext} 
\eea 
The first term is the ``standard" 
chiral symmetry breaking term 
in the linear sigma model,
while the second term arises
naturally in $\chi$PT. 
The third term proportional to $\varepsilon_3 $ 
was discussed in, e.g. \cite{c79,bc79,dm00},\
and we remark that a term proportional 
to $\bar\psi \psi$ appears in $\chi$PT 
with a coefficient proportional to $m_\pi^2 c_1$,
where $c_1$ is a low-energy constant
in $\chi$PT \cite{gss88}.

\vspace{3mm} 

As usual, we redefine the scalar field relative 
to its vacuum expectation value, $<\!\sigma\! >_0 = f_\pi$,
and introduce the new scalar field $s$ defined by 
$s = \sigma - f_\pi$. 
The requirement that the energy is minimum for 
$<\!\sigma \!>_0 = f_\pi$ 
gives the following relation
$\mu_0^2 - \lambda f^2_\pi = 
- \varepsilon_1 / f_\pi $\,. 
The pion mass is found to be 
\bea 
m_\pi^2 = \varepsilon_1 / f_\pi  \; 
+\; 2\, \varepsilon_2 \; .
\eea
In what follows
we evaluate the $\pi N$
scattering amplitude using the lagrangian in 
Eq.(\ref{eq:Lsigma}) properly modified to 
account for the redefinition of the 
scalar field,  $\sigma$ $\to$ $s$, explained above.

\vspace{3mm} 

The $\pi N$ scattering $T$-matrix is conventionally 
written as 
\bea
T_{\alpha \beta} &=& T^{(+)} \; \delta_{\alpha \beta} 
+ T^{(-)} \; \frac{1}{2} [\tau_\alpha , \tau_\beta ]
\label{eq:Tab}
\eea
where $\alpha$ and $\beta$ are the initial and 
final pion isospin indices, respectively,
and $T^{(\pm)} $ are defined as  
\bea 
T^{(\pm)}=A^{(\pm)}+B^{(\pm)}
\frac{1}{2} \gamma_\mu (k_1^\mu +k_2^\mu )
\label{eq:dm00}
\eea
Here $k_1$ and $k_2$ are 
the incoming and outgoing pion momenta,
respectively, in the center-of-mass system.
It is understood that,
in order to obtain the scattering amplitude,
$T_{\alpha \beta}$ should be sandwiched 
between the relevant Dirac spinors 
(which however are suppressed in Eq.(\ref{eq:Tab})).
To compare the $\pi N$ amplitudes evaluated 
in ELSM with the ones obtained in HB$\chi$PT, 
we have to treat the nucleons in ELSM 
as heavy, non-relativistic  
fields of mass $m_N$. 
Therefore, the ELSM amplitudes, $A^{(\pm)}$ and $B^{(\pm)}$, 
in Eq.(\ref{eq:dm00}) 
and the Dirac spinors
describing the initial and final nucleons in ELSM 
need to be expanded in powers of 
$1/m_N \equiv 1/M $. 
The corresponding 
non-relativistic $\pi N$ scattering amplitudes,  
$g^{(\pm)}$ and $h^{(\pm)}$, are customarily  
defined by
\bea
\tilde{T}_{\alpha \beta }&=& \Big[ g^{(+)} +
i\vec{\sigma}\cdot (\vec{k}_1\times \vec{k}_2)  \; 
 h^{(+)} \Big] \; \delta_{\alpha \beta} 
 + \Big[ g^{(-)} +
i\vec{\sigma}\cdot (\vec{k}_1\times \vec{k}_2)  
\; h^{(-)} \Big] \; \frac{1}{2} 
[\tau_\alpha,\; \tau_\beta] \Big] \; . 
\label{eq:Tnonrel} 
\eea
It is understood here
that $\tilde{T}_{\alpha,\beta}$ 
is to be sandwiched between the initial and final 
nucleon Pauli spinors and iso-spinors 
to yield the scattering amplitude. 
The amplitudes, $ g^{(\pm)}$ and $ h^{(\pm)}$,
calculated in ELSM and HB$\chi$PT
are denoted by $ g_{ESM}^{(\pm)}$, $ h_{ESM}^{(\pm)}$,
$ g_{\chi PT}^{(\pm)}$ and $ h_{\chi PT}^{(\pm)}$,
respectively.
Comparison between $ g_{ESM}^{(\pm)}$ and 
$ g_{\chi PT}^{(\pm)}$ and between
$ h_{ESM}^{(\pm)}$ and $ h_{\chi PT}^{(\pm)}$
is our main concern in what follows.

\vspace{3mm}
In Ref. \cite{dm00} the elastic $\pi N$ 
scattering amplitude in ELSM
was calculated in the tree approximation and the 
expressions for the four amplitudes, 
$A^{(+)}$, $A^{(-)}$, $B^{(+)}$ and $B^{(-)}$,  
are given in Eqs.(17a-d) 
of Ref.\cite{dm00}.\footnote{We remark that 
the overall sign of 
the amplitude $B^{(-)}$, Eq.(17d) in \cite{dm00},  
should be changed.}
The corresponding 
$ g_{ESM}^{(\pm)}$ and $ h_{ESM}^{(\pm)}$ 
amplitudes were derived in Ref.\cite{wpa02}.
The $\pi N$ scattering amplitude in HB$\chi$PT 
was evaluated by, e.g., Meissner et al.\cite{meissner}. 
For our present purposes,
we only need the tree approximation amplitudes. 
The amplitudes, 
$ g_{\chi PT}^{(\pm)}$ and $ h_{\chi PT}^{(\pm)}$,
were rederived in Ref.\cite{wpa02} and 
it has been confirmed that 
in the tree approximation 
the results agree with those of Ref.\cite{meissner}. 

\vspace{3mm}
As mentioned earlier, Weinberg's 
non-linear sigma model 
can be derived from ELSM
in the limit of $m_\sigma \to \infty$. 
Therefore, to facilitate comparison 
with the HB$\chi$PT expressions, the 
amplitudes obtained in ELSM are further simplified
by assuming that  
$m_\sigma$ is heavy compared to the pion mass and energy
and expanding the amplitudes in powers of $1/m_\sigma$. 
We assume that $m_\sigma\simeq$ $M\simeq$ 
$\Lambda_\chi \simeq 1$ GeV, whereas 
$m_\pi , \; \omega$ and $\sqrt{t}$ are of order 
$Q \ll  \Lambda_\chi $. 
We also assume that the chiral symmetry breaking parameters 
$\varepsilon_i$ are of order $Q^2$.
We restrict our comparison to  
the lowest powers of $Q$ in each amplitude, and we  
use the fact that the 
LECs, $c_i$ ($i=1, 2, 3$), in HB$\chi$PT are 
of the natural order of magnitude 
($c_i\Lambda_\chi\sim 1$). 

\vspace{3mm} 

In comparing the amplitudes obtained in the
two approaches under consideration, 
we find it convenient to introduce
the following decompositions:
\bea 
g_{ESM}^{(\pm)} = \tilde{g}^{(\pm)} 
+ \delta g_{ESM}^{(\pm)} \;, 
\;\;\;\;\; 
g_{\chi PT }^{(\pm)} = \tilde{g}^{(\pm)} 
+ \delta g_{\chi PT }^{(\pm)} \;,
\eea
\bea
h_{ESM}^{(\pm)} = \tilde{h}^{(\pm)} 
+ \delta h_{ESM}^{(\pm)} \;,
\;\;\;\;\; 
h_{\chi PT }^{(\pm)} = \tilde{h}^{(\pm)} 
+ \delta h_{\chi PT }^{(\pm)} \;.
\eea 
In the above,
$\tilde{g}^{(\pm)}$ represents
the part that has a common analytic expression
between $g_{ESM}^{(\pm)}$  \cite{dm00,wpa02} 
and $g_{\chi PT }^{(\pm)}$ \cite{wpa02,meissner}, 
whereas $\delta g_{ESM}^{(\pm)}$ and 
$\delta g_{\chi PT }^{(\pm)}$ represent
the parts that do not have common analytic expressions.
Similarly for $\tilde{h}^{(\pm)}$. 
The terms common between ELSM and HB$\chi$PT
are given by
\bea
\tilde{g}^{(+)} &=& 
\left(\frac{g_A^2}{f_\pi^2}\right) 
\left( \frac{ 
2\omega^2m_\pi^2-\omega^4+(\vec{k}_1\cdot\vec{k}_2)^2 }{4M\omega^2}
\right) 
\nonumber \\ 
\tilde{g}^{(-)} &=& \left(\frac{\omega}{2f_\pi^2} \right) 
-\left(\frac{g_A^2}{f_\pi^2}\right)
\frac{ \vec{k}_1\cdot\vec{k}_2 }{2\omega} 
+\frac{1}{f_\pi^2}
\left( 
\frac{ 
\omega^4-m_\pi^2\omega^2+\omega^2 (\vec{k}_1\cdot\vec{k}_2) 
}{4M\omega^2} \right) 
\nonumber \\ &+& 
\left(\frac{g_A^2}{f_\pi^2}\right) 
\frac{1}{4M\omega^2}\;
\Big[ -2\omega^4+2\omega^2m_\pi^2-m_\pi^2(\vec{k}_1\cdot\vec{k}_2)
-\omega^2 (\vec{k}_1\cdot\vec{k}_2) +(\vec{k}_1\cdot\vec{k}_2)^2 
\Big] 
\nonumber \\
\tilde{h}^{(+)} &=&
-\frac{g_A^2}{2\omega f_\pi^2} - \left(\frac{g_A^2}{f_\pi^2}\right)
\left( \frac{ 
\omega^2+m_\pi^2-\vec{k}_1\cdot\vec{k}_2 }{4M\omega^2}  \right)
\nonumber \\ 
\tilde{h}^{(-)} &=& \frac{1}{f_\pi^2} \left(
\frac{ \omega^2 - g_A^2 (\vec{k}_1\cdot\vec{k}_2)  }{
4M\omega^2} \right) 
\eea

\vspace{3mm}

As for $\delta g^{(\pm )}$ and 
$\delta h^{(\pm )}$,
our ELSM calculation leads to the following results: 
\bea
\delta g_{ESM}^{(+)} &=& \frac{M}{f_\pi^2} 
\Big\{
-
\frac{  \Big[ t-m_\pi^2 +2\varepsilon_2 \Big] }{ m_\sigma^2} 
-\frac{\varepsilon_3}{M}  
+ {\cal O}(M^{-2},  m_\sigma^{-2}) 
\Big\}
\label{eq:gplus1} 
\\ 
\delta g_{ESM}^{(-)} &=& \frac{g_A}{f_\pi^2} \Big\{
\frac{\varepsilon_3  
\left( t-2m_\pi^2 \right) }{2 M\omega} +\cdots 
\Big\}
\label{eq:gminus1} 
\\ 
\delta h_{ESM}^{(+)} &=& -\frac{1}{f_\pi^2} 
  \left( 
\frac{t-m_\pi^2 
+ 2 \varepsilon_2 }
{4Mm_\sigma^2} - \frac{g_A\varepsilon_3}{M \omega } 
\right)   +\cdots 
\label{eq:hplus1}
\\ 
\delta h_{ESM}^{(-)} &=& 0 .  
\label{eq:hminus1}
\eea 
Meanwhile, an HB$\chi$PT calculation (in tree approximation)
gives, to the order $Q^2/\Lambda_\chi^2$,
the following results.
\bea
\delta g_{\chi PT}^{(+)} &=& \left(
c_3 \;\Big[t-2m_\pi^2\Big]+4m_\pi^2\; c_1 -2\omega^2\; c_2 
\right)\frac{1}{f_\pi^2}
\label{eq:gplus2}
\\ 
\delta g_{\chi PT}^{(-)} &=& 0 
\label{eq:gminus2}
\\ 
\delta h_{\chi PT}^{(+)} &=& 0 
\label{eq:hplus2}
\\
\delta h_{\chi PT}^{(-)} &=& -\frac{c_4}{f_\pi^2} ,  
\label{eq:hminus2}
\eea 
where $t$ = $(k_1-k_2)^2$ is 
the four-momentum transfer~\cite{wpa02}.

\vspace{3mm}

We now discuss to what extent the sigma model 
(ELSM in our case) 
simulates the effective field theory (here HB$\chi$PT).
We first look at the results for $g^{(+)}$.
It is informative to examine
what values ELSM gives   
to the LECs, $c_i$ 
($i = 1, \cdots , 4$), that appear in HB$\chi$PT.  
To this end, let us impose the requirement 
$\delta g_{ESM}^{(+)} = 
\delta g_{\chi PT}^{(+)} $.  
Comparison of the momentum transfer ($t$) 
and energy ($\omega$) 
dependences in Eq.(\ref{eq:gplus1}) 
and Eq.(\ref{eq:gplus2}) leads us to identify 
\bea 
c_3 \sim - \frac{M}{m_\sigma^2}, \; \; \; 
c_1 \sim - \frac{M}{m_\sigma^2}\left(
\frac{m_\pi^2 +2\varepsilon_2}{4m_\pi^2}\right) -
\frac{\varepsilon_3}{4m_\pi^2} 
\; \; \; {\rm and} \; \; \; 
c_2=0 
\; . 
\label{eq:LECs}
\eea 
The result $c_2=0$ means that  
ELSM fails to generate 
the energy-dependence of
$g^{(+)}$ required by HB$\chi$PT. 
With the use of the sigma-meson mass scale,
$m_\sigma \sim M \sim  1 \; $GeV, 
we find
$c_3 \sim c_1 \sim - 1\;$GeV$^{-1}$. 
These results are not inconsistent with
those found in Ref.~\cite{bkm97} based on  
the resonance saturation assumption.  
There it was shown that the $\Delta$-resonance gives a 
major contribution to $c_2$ and $c_3$,  
whereas the empirical value of $c_1$ can be explained by 
a scalar resonance contribution in the two-pion channel. 
Furthermore, this scalar-meson resonance 
was found to give a $\sim$30\% contribution
to $c_3$~\cite{bkm97}. 
These features are compatible with our finding
that ELSM, which contains no $\Delta$-field, 
leads to $c_2=0$ and to the value of $c_3$
that is significantly smaller than 
the empirically determined value. 
As for $g^{(-)}$,
we notice that 
$\delta g_{ESM}^{(-)}$ in Eq.(\ref{eq:gminus1}) is of order 
${\cal O}(Q^3)$, i.e., this amplitude has no terms of order 
${\cal O}(Q^2)$.  This feature is consistent with 
Eq.(\ref{eq:gminus2}). 

\vspace{3mm} 
Regarding the ``spin-flip" amplitudes 
$h^{(\pm)}$, we note that $h^{(\pm)}$
in Eq.(\ref{eq:Tnonrel}) are accompanied
by a factor of ${\cal O}(Q^2)$.
This means that, to the chiral order 
under consideration, the comparison of
the ELSM and HB$\chi$PT results for $h^{(\pm)}$
should be limited to the ${\cal O}(1)$ terms.
Comparison between 
$\delta h_{ESM}^{(-)}$ in Eq.(\ref{eq:hminus1}) and
$\delta h_{\chi PT}^{(-)}$ in Eq.(\ref{eq:hminus2})
leads to the conclusion that $c_4=0$. 
This implies that ELSM cannot generate the iso-vector, 
spin-dependent term in the $\pi N$ scattering amplitude 
predicted by HB$\chi$PT. 
Again we refer to Ref.~\cite{bkm97}, 
where it is shown that 
the empirical value of $c_4$ can 
be explained, within the resonance saturation 
assumption, 
by dominant contributions from the $\Delta$-resonance
and the $\rho$-meson. 
Since none of these hadrons are included in ELSM, 
it should come as no surprise
that $c_4=0$ in ELSM. 
As for $\delta h^{(+)}$, 
Eq.(\ref{eq:hplus1}) indicates that 
$\delta h^{(+)}_{ESM}$ is of ${\cal O}(Q)$,
{\it i.e.,} it has no contribution of ${\cal O}(1)$.
This feature is consistent with the fact
that HB$\chi$PT generates no $\delta h^{(+)}$ amplitude
of chiral orders lower than $Q^2$
[see Eq.(\ref{eq:hplus2})].

\vspace{3mm} 
We remark en passant that,
if we take the limit 
$m_\sigma \to \infty$ \underline{and} 
require $\varepsilon_3 = 0$,
then ELSM leads to 
$c_1=c_2=c_3=c_4=0$,
and ELSM and HB$\chi$PT give 
identical tree approximation 
$\pi N$ scattering amplitudes, 
$g_{ESM}^{(\pm)} = g_{\chi PT}^{(\pm)} 
= \tilde{g}^{(\pm)}$, and 
$h_{ESM}^{(\pm)} = h_{\chi PT}^{(\pm)} 
= \tilde{h}^{(\pm)}$.  
This is however a very special case.

\vspace{3mm}

The above comparison indicates that, 
in general, the 
ELSM fails to reproduce some of 
the $\pi N$ scattering amplitude properties,
(e.g., the energy dependence)
that are predicted by HB$\chi$PT. 
It is well known that $\Delta$ degrees of freedom 
play a role in describing $\pi N$ scattering 
even at very low energies.
In HB$\chi$PT, although only 
the pion and nucleon are explicit degrees of freedom,
the effects of the $\Delta$-resonance
are subsumed in the LECs, $c_2, c_3$ and $c_4$. 
By contrast, 
the $\Delta$-field is normally not included 
in sigma models.
This difference seems to be the main cause of
the failure for ELSM to reproduce 
the HB$\chi$PT results.
A lesson we learn from the present study
is that, although the linear sigma model
(either in its original version
or in the form of ELSM) is often used
as a convenient tool
for exploring consequences of chiral symmetry 
in nuclear physics,  
conclusions obtained in such studies
should be taken with caution.

\vskip 3mm 

W.P.A thanks T.-S. Park for many useful suggestions. 
We are grateful
to the referee for his/her extremely useful
comments and for drawing our attention
to the results given in Ref.\cite{bkm97}. 
This work is supported in part by the 
US NSF Grant No. PHY-0140214.


\vskip 3mm \noindent

\begin{thebibliography}{99} 
\bibitem{GML60} 
M. Gell-Mann and M. Levy, 
Nuovo Cimento, {\bf 16}, 705 (1960)

\bibitem{ericson} 
J. Delorme, G. Chanfray and M. Ericson,
Nucl. Phys. A, {\bf 603}, 239 (1996).

\bibitem{bjn68}
J.D. Bjorken and M. Nauenberg, 
Ann. Rev. Nucl. Sci. {\bf 18}, 299 (1968).

\bibitem{lee68}
H.B. Lee, {\it Chiral Dynamics} 
(Gordon and Breach, New York, 1972), Chap.10.c. 

\bibitem{dm00}
V. Dmitrasinovic and F. Myhrer, 
Phys. Rev. C, {\bf 61}, 025205 (2000) 

\bibitem{weinberg}
S. Weinberg, 
Physica A, {\bf 96}, 327 (1979).

\bibitem{wei90}
S. Weinberg, 
Phys. Lett. B, {\bf 251}, 288 (1990); 
{\it ibid}, {\bf 295}, 114 (1992).

\bibitem{gss88} 
J. Gasser, M. Sainio and  S. Svarc, 
Nucl. Phys. B, {\bf 307}, 779 (1988).

\bibitem{jm91}
E. Jenkins and A. Manohar,
Phys. Lett. B, {\bf 255}, 558 (1991).

\bibitem{vanK99}
U. van Kolck, Prog. Part. Nucl. Phys. {\bf 43}, 337 (1999). 

\bibitem{bkm95}
V. Bernard, N. Kaiser and U.-G. Meissner,
Int. J. Mod. Phys. E, {\bf 4}, 193 (1995).


\bibitem{comment1}
In this connection, it is worth noting 
that Weinberg's non-linear chiral lagrangian 
is the lowest order lagrangian of $\chi$PT, 
see, e.g. Ref.~\cite{gss88}.

\bibitem{c79} 
D.K. Campbell, 

Phys. Rev. C, {\bf 19}, 1965 (1979)

\bibitem{bc79} 
G. Baym and D.K. Campbell, 
in {\it Mesons in Nuclei}, 
eds. M. Rho and D.H. Wilkinson 
(North Holland, Amsterdam, 1979), p. 1033. 

\bibitem{wpa02} 
W.P. Alvarez, Ph.D. thesis, 
University of South Carolina (2002), unpublished. 

\bibitem{meissner} 
N. Fettes et al.,
Nucl. Phys. A, {\bf 640}, 199 (1998).

\bibitem{bkm97} V. Bernard, N. Kaiser and U.-G. Meissner,
Nucl. Phys. {\bf A 615}, 483 (1997). 


\end{thebibliography}
\end{document}